\documentclass[a4paper,runningheads]{llncs}

\usepackage[latin1]{inputenc}

\usepackage{amsmath}
\usepackage{amssymb}

\usepackage[table]{xcolor}
\usepackage{graphicx}
\usepackage{tikz}
\usetikzlibrary{positioning}

\usepackage[bookmarks,bookmarksopen,bookmarksdepth=2,pdftex,colorlinks,urlcolor=red!50!black,linkcolor=red!50!black,citecolor=red!50!black]{hyperref}
\usepackage{cleveref}
\usepackage[numbers,sort]{natbib}

\usepackage[ruled,vlined,nofillcomment,linesnumbered]{algorithm2e}
\crefname{algocf}{alg.}{algs.}
\Crefname{algocf}{Alg.}{Algorithms}

\Crefname{figure}{Fig.}{Figs.}
\Crefname{table}{Tab.}{Tabs.}

\usepackage{multirow}

\usepackage{multicol}

\usepackage{url}

\newcommand{\pivmds}{PivotMDS}
\newcommand{\bp}{1-stress}
\newcommand{\stress}{full stress}
\newcommand{\sparsefif}{sparse 50}
\newcommand{\sparseone}{sparse 100}
\newcommand{\sparsetwo}{sparse 200}
\newcommand{\maxent}{maxent}
\newcommand{\grip}{GRIP}
\newcommand{\mars}{MARS}
\newcommand{\marsone}{\mars~100}
\newcommand{\marstwo}{\mars~200}

\newcommand{\addDistribution}[1]{\includegraphics[width=9mm,height=3mm]{#1}}
\newcommand{\addDegreeDist}[1]{\addDistribution{#1/degree.pdf}}
\newcommand{\addDistanceDist}[1]{\addDistribution{#1/distance_dist.pdf}}

\newcommand{\addGraphicFormat}[2]{\resizebox{40mm}{40mm}{\includegraphics[]{#1#2}}}

\newcommand{\addPNG}[1]{\addGraphicFormat{#1}{.png}}

\newcommand{\addLayoutToTableComb}[1]{
\multirow{2}{*}{\rotatebox[origin=c]{90}{\LARGE{\textbf{#1}}}}
& \addPNG{#1/stress_layout}
& \addPNG{#1/sparse_200_layout}
& \addPNG{#1/sparse_100_layout}
& \addPNG{#1/sparse_50_layout}
& \addPNG{#1/maxent_layout}
& \addPNG{#1/mars_200_layout}
& \addPNG{#1/mars_100_layout}
& \addPNG{#1/grip_layout}
& \addPNG{#1/bp_layout}
& \addPNG{#1/pivmds_layout}}

\newcommand{\addLayoutAndOverviewToTable}[1]{\addLayoutToTableComb{#1} \\\addOverviewToTableComb{#1}}

\newcommand{\dimension}{\ensuremath{\alpha}}
\newcommand{\layVar}{\ensuremath{x}}
\renewcommand{\O}[1]{\ensuremath{\mathcal{O}(#1)}}
\renewcommand{\P}{\ensuremath{\mathcal{P}}}
\newcommand{\R}{\ensuremath{\mathcal{R}}}

\DeclareMathOperator*{\argmax}{arg\,max}

\graphicspath{{figs/}}

\begin{document}

\title{A Sparse Stress Model\thanks{We gratefully acknowledge financial support from Deutsche Forschungsgemeinschaft under grant Br~2158/11-1}}
\titlerunning{A Sparse Stress Model}
\author{Mark Ortmann, Mirza Klimenta \and Ulrik Brandes}
\authorrunning{M.~Ortmann, M.~Klimenta and U.~Brandes}
\tocauthor{M.~Ortmann, M.~Klimenta, and U.~Brandes}
\toctitle{A Sparse Stress Model}
\institute{Computer \& Information Science, University of Konstanz \email{Mark.Ortmann@uni-konstanz.de}}

\maketitle

\begin{abstract}
Force-directed layout methods constitute the most common approach to draw general graphs.
Among them, stress minimization produces layouts of comparatively high quality but also imposes comparatively high computational demands.
We propose a speed-up method based on the aggregation of terms in the objective function.
It is akin to aggregate repulsion from far-away nodes during spring embedding
but transfers the idea from the layout space into a preprocessing phase.
An initial experimental study informs a method to select representatives, 
and subsequent more extensive experiments indicate
that our method yields better approximations of minimum-stress layouts in less time than related methods.
\end{abstract}

\section{Introduction}

There are two main variants of force-directed layout methods,
expressed either in terms of forces to balance or an energy function to minimize~\cite{b-dpa-01,k-fdda-13}.
For convenience, we refer to the former as spring embedders and to the latter as multidimensional scaling~(MDS) methods.

Force-directed layout methods are in wide-spread use and of high practical significance, but their scalability is a recurring issue.
Besides investigations into adaptation, robustness, and flexibility,
much research has therefore been devoted to speed-up methods~\cite{yl-vlg-15}.
These efforts address, e.g., the speed of convergence~\cite{fr-gdfdp-91,flm-fal-95} or the time per iteration~\cite{bh-nln-86,g-mp-87}.
Generally speaking, the most scalable methods are based on multi-level techniques~\cite{w-mfd-03,gk-grip-00,hj-fm3-04,hu-hqg-05}.

Experiments~\cite{bp-esdbgd-09} suggest that minimization of the stress function~\cite{g-maed-66}
\begin{equation} \label{eq:stress}
	s(\layVar) = \sum_{i<j}w_{ij}(||\layVar_i-\layVar_j|| - d_{ij})^2
\end{equation}
is the primary candidate for high-quality force-directed layouts $\layVar\in(\mathbb{R}^{2})^V$
of a simple undirected graph $G=(V,E)$ with $V=\{1,\ldots,n\}$ and $m=|E|$.
The target distances $d_{ij}$ are usually chosen to be the graph-theoretic distances, the weights set to $w_{ij} = 1/d_{ij}^2$, 
and the dominant method for minimization is majorization~\cite{gkn-sm-04}.
Several variant methods reduce the cost of evaluating the stress function
by involving only a subset of node pairs over the course of the algorithm~\cite{c-cp-97,gk-grip-00,bsww-tpsmm-02}.
If long distances are represented well already, for instance because of initialization with a fast companion algorithm,
it has been suggested that one restrict further attention to short-range influences from $k$-neighborhoods only~\cite{bp-esdbgd-09}.

We here propose to stabilize the sparse stress function restricted to $1$-neigh\-bor\-hoods~\cite{bp-esdbgd-09}
with aggregated long-range influences inspired by the use of Barnes \& Hut approximation~\cite{bh-nln-86}
in spring embedders~\cite{t-jiggle-98}.
Extensive experiments suggest how to determine representatives for individually weak influences,
and that the resulting method represents a favorable compromise between efficiency and quality.

Related work is discussed in more detail in the next section.
Our approach is derived in \Cref{sec:ssm}, and evaluated in \Cref{sec:ed}.
We conclude in \Cref{sec:con}.

\section{Related Work}\label{sec:rel}

While we are interested in approximating the full stress model of Eq.~\eqref{eq:stress},
there are other approaches capable of dealing with given target distances
such as the strain model~\cite{bp-pmds-07,kb-cmdsnp-12,st-gvl-02} or the Laplacian~\cite{kch-ace-02, h-qpa-1970}.

An early attempt to make the full stress model scale to large graphs is \grip~\cite{gk-grip-00}.
Via a greedy maximal independent node set filtration, this multi-level approach constructs a hierarchy of more and more coarse graphs. While a sparse stress model calculates the layout of the coarsened levels, the finest level is drawn by a localized spring-embedder~\cite{fr-gdfdp-91}. Given the coarsening hierarchy for graphs of bounded degree, \grip~requires \O{nk^2} time and \O{nk} space with $k = \log \max\{d_{ij} : i,j \in V\}$.

Another notable attempt has been made by Gansner et al.~\cite{ghn-msm-13}. Like the spring embedder the maxent-model is split into two terms:
\begin{equation*}
 \sum_{\{i,j\} \in E} w_{ij}(||\layVar_i-\layVar_j|| - d_{ij})^2  - \alpha \sum_{\{i,j\} \not \in E} \log ||\layVar_i-\layVar_j||
\end{equation*}
The first part is the \bp~model~\cite{gk-grip-00, bp-pmds-07}, while the second term tries to maximize the entropy. Applying Barnes \& Hut approximation technique~\cite{bh-nln-86}, the running time of the maxent-model can be reduced from \O{n^2} per iteration to \O{m+n\log n}, e.g., using quad-trees~\cite{q-vca-00,t-ggd-99}. In order to make the maxent-model even more scalable Meyerhenke et al.~\cite{mns-mms-15} embed it into a multi-level framework, where the coarsening hierarchy is constructed using an adapted size-constrained label propagation algorithm.

Gansner et al.~\cite{ghk-coast-13}, inspired by the idea of decomposing the stress model into two parts, proposed COAST. The main difference between COAST and \maxent~is that it adds a square to the two terms in the \bp~part and that the second term is quadratic instead of logarithmic. Transforming the energy system of COAST allows one to apply fast-convex optimization techniques making its running time comparable to the \maxent~model.

While all these approaches somewhat steer away from the stress model, \mars~\cite{khks-mars-12} tries to approximate the solution of the full stress model. Building on a result of Drineas et al.~\cite{dfkvv-svd-04}, \mars~requires only $k \ll n$ instead of $n$ single-source shortest path computations. Reconstructing the distance matrix from two smaller matrices and by setting $w_{ij} = 1/d_{ij}$, \mars~runs in \O{kn+n\log n+ m} per iteration with a preprocessing time in \O{k^3 + k(m+n \log n + k^2n)}, and a space requirement in \O{nk}. 

\section{Sparse Stress Model}\label{sec:ssm}
The full stress model, Eq.~\eqref{eq:stress}, is in our opinion the best choice to draw general graphs, not least because of its very natural definition. However, its \O{n^2} running time per iteration and space requirement, and expensive processing time of \O{n(m+n\log n)}, hamper its way into practice. 

The reason sparse stress models are still in early stages of development is that the adaption to large graphs requires not just a reduction in the running time per iteration, but also the preprocessing time and its associated space requirement. Where these problems originate from is best explained by rewriting Eq.~\eqref{eq:stress} to the following form:
\begin{equation}\label{eq:twoStress}
	s(\layVar) = \sum_{\{i,j\} \in E} w_{ij}(||\layVar_i-\layVar_j|| - d_{ij})^2 + \sum_{\{i,j\} \in {V \choose 2} \setminus E} w_{ij}(||\layVar_i-\layVar_j|| - d_{ij})^2
\end{equation}
As minimizing the first term only requires \O{m} computations and all $d_{ij}$ are part of the input, solving this part of the stress model can be done efficiently. Yet, the second term requires an all-pairs shortest path computation (APSP), \O{n^2} time per iteration, and in order to stay within this bound \O{n^2} additional space. We note that the \bp~approaches presented in \Cref{sec:rel} of Gajer et al.~\cite{gk-grip-00} and Brandes \& Pich~\cite{bp-pmds-07} ignore the second term, while Gansner et al.~\cite{ghk-coast-13,ghn-msm-13} replace it. Discounting the problems arising from the APSP computation, we can see that the spring embedder suffered from exactly the same problem, namely the computation of the second term -- there called repulsive forces. Barnes \& Hut introduced a simple, yet ingenious and efficient solution, namely to approximate the second term by using only a subset of its addends. 

To approximate the repulsive forces operating on node $i$ Barnes \& Hut partition the graph. Associated with each of these $\mathcal{O}(\log n)$ partitions is an artificial representative, a so called super-node, used to approximate the repulsive forces of the nodes in its partition affecting $i$. However, as these super-nodes have only positions in the euclidean space, but no graph-theoretic distance to any node in the graph they cannot be processed in the stress model. Furthermore, deriving a distance for a super-node as a function of the graph-theoretic distance of the nodes it represents would require an APSP computation, which is too costly, and since the partitioning is computed in the layout space, probably not a good approximation. Choosing a node from the partition as a super-node would not solve the problems, not least because the partitioning changes over time.

Therefore, adapting this approach cannot be done in a straightforward manner. However, the model we are proposing sticks to its main ideas. In order to reduce the complexity of the second term in Eq.~\eqref{eq:twoStress}, we restrict the stress computation of each $i \in V$ to a subset $\P \subseteq V$ of $k = |\P|$ representatives, from now on called pivots. The resulting sparse stress model, where $N(i)$ are the neighbors of $i$ and $w_{ip}'$ are adapted weights, has the following form:
\begin{equation}\label{eq:sparse}
	s'(\layVar) = \sum_{\{i,j\} \in E} w_{ij}(||\layVar_i-\layVar_j|| - d_{ij})^2 + \sum_{i \in V}\sum_{p \in \P \setminus N(i)}  w'_{ip}(||\layVar_i-\layVar_p|| - d_{ip})^2
\end{equation}
Note that GLINT~\cite{im-glint-12} uses a similar function, yet the pivots change in each iteration, no weights are involved, and it is assumed that $d_{ip}$ is accessible in constant time.

Just like Barnes \& Hut, we associate with each pivot $p \in \P$ a set of nodes $\R(p) \subseteq V$, where $p \in \R(p)$, $\bigcup_{p \in \P} \R(p) = V$, and $\R(p) \cap \R(p') = \emptyset$ for $p,p' \in \P$. However, we propose to use only one global partitioning of the graph that does not change over time. Still, just like the super-nodes, we want that the pivots are representative for their associated region. In terms of the localized stress minimization algorithm~\cite{gkn-sm-04} this means we want that for each $i \in V$ and $p \in \P$
\begin{equation*}\label{eq:lma}
	\frac{\sum_{j \in \R(p)} w_{ij} (\layVar_j^{\dimension} + d_{ij}(\layVar_i^{\dimension} - \layVar_j^{\dimension}) / ||\layVar_i-\layVar_j||)}{\sum_{j \in \R(p)}w_{ij}} \approx \layVar_p^{\dimension} + \frac{d_{ip}(\layVar_i^{\dimension} - \layVar_p^{\dimension})}{||\layVar_i-\layVar_p||},
\end{equation*}
where $\dimension$ is the dimension. As the left part is the weighted average of all positional votes of $j \in \R(p)$ for the new position of $i$, we require $p$ to fulfill the following requirements in order to be a good representative:
\begin{itemize}
	\item The graph-theoretic distances to $i$ from all $j \in \R(p)$ should be similar to $d_{ip}$
	\item The positions of $j \in \R(p)$ in $\layVar$ should be well distributed in close proximity around $p$.
\end{itemize}
We propose to construct the partitioning induced by $\R$ only based on the graph structure, not on the layout space, and associate each node $v \in V$ with $\R(p)$ of the closest pivot subject to their graph-theoretic distance. As our algorithm incrementally constructs $\R$, ties are broken by favoring the currently smallest partition. Given the case that $\P$ has been chosen properly and since all nodes in $\R(p)$ are at least as close to $p$ as to any other pivot, and consequently in the stress drawing, it is appropriate to assume that both conditions are met.

\begin{algorithm}[tb]
\DontPrintSemicolon
\KwIn{Graph $G=(V,E)$ with $w: E \rightarrow \mathbb{R}_{>0}$, and $k$ number of pivots.}
\KwOut{$\dimension-$dimensional layout $\layVar \in (\mathbb{R}^{\dimension})^V$ }
sample \P~with $|\P| = k$\;
calculate $\R$, all adapted weights $w_{ip}'$, and all $d_{ip}$ via weighted MSSP\;
$\layVar \leftarrow $\texttt{PivotMDS(G)}~\cite{bp-pmds-07}\;
rescale $\layVar$ such that $\sum_{\{i,j\}\in E}||\layVar_i-\layVar_j|| = \sum_{\{i,j\}\in E}w_{ij}$\;
\While{relative positional change $> 10^{-4}$}{
	\ForEach{$i \in V$}{
		\ForEach{dimension \dimension}{
			$t^{\dimension} \leftarrow \frac{\sum_{j \in N(i)} w_{ij} \left(\layVar_j^{\dimension} + \frac{d_{ij}(\layVar_i^{\dimension} - \layVar_j^{\dimension})}{||\layVar_i-\layVar_j||}
				\right) + \sum_{p \in \P \setminus N(i)} w_{ip}' \left(\layVar_p^{\dimension} + \frac{d_{ip}(\layVar_i^{\dimension} - \layVar_p^{\dimension})}{||\layVar_i-\layVar_p||}\right)}{\sum_{j \in N(i)}w_{ij}+ \sum_{p \in \P \setminus N(i)}w_{ij}'}$
		}
		$\layVar_i \leftarrow t$\;
	}
}
\caption{Sparse Stress}
\label{alg:sparse}
\end{algorithm}

Even if the positional vote of each pivot is optimal w.r.t.~$\R(p)$, it is still not enough to approximate the full stress model. In the full stress model the iterative algorithm to minimize the stress moves one node at a time while fixing the rest. By setting node $i$'s position in dimension $\dimension$ to
\begin{equation*}
	\layVar_i^{\dimension} = \frac{\sum_{j \not= i} w_{ij} (\layVar_j^{\dimension} + d_{ij}(\layVar_i^{\dimension} - \layVar_j^{\dimension}) / ||\layVar_i-\layVar_j||)}{\sum_{j \not= i}w_{ij}},
\end{equation*}
it can be shown that the stress monotonically decreases~\cite{gkn-sm-04}. However, in our model we move node $i$ according to
\begin{equation}\label{eq:ssmin}
	\layVar_i^{\dimension} = \frac{\sum_{j \in N(i)} w_{ij} \left(\layVar_j^{\dimension} + \frac{d_{ij}(\layVar_i^{\dimension} - \layVar_j^{\dimension})}{||\layVar_i-\layVar_j||}
	\right) + \sum_{p \in \P \setminus N(i)} w_{ip}' \left(\layVar_p^{\dimension} + \frac{d_{ip}(\layVar_i^{\dimension} - \layVar_p^{\dimension})}{||\layVar_i-\layVar_p||}\right)}{\sum_{j \in N(i)}w_{ij}+ \sum_{p \in \P \setminus N(i)}w_{ij}'}.
\end{equation}
This implies that in order to find the globally optimal position of $i$ we furthermore have to find weights $w_{ip}'$, such that $\frac{w'_{ip}}{\sum_{j\in N(i)} w_{ij} + \sum_{p \in \P\setminus N(i)}  w'_{ip}} \approx \frac{\sum_{j\in \R(p)} w_{ij}}{\sum_{i \not= j}w_{ij}}$.
Since our goal is only to reconstruct the proportions, and our model only knows the shortest-path distance between all nodes $i \in V$ and $p \in \P$, we set $w_{ip}' = s/d_{ip}^2$ where $s \geq 1$. At the first glance setting $s = |\R(p)|$ seems appropriate, since $p$ represents $|\R(p)|$ addends of the stress model. Nevertheless, this strongly overestimates the weight of close partitions. Therefore, we propose to set $s = |\{j \in \R(p) : d_{jp} \leq d_{ip} /2\}|$. This follows the idea that $p$ is only a good representative for the nodes in $\R(p)$ that are at least as close to $p$ as to $i$. Since the graph-theoretic distance between $i$ and $j \in \R(p)$ is unknown, our best guess is that $j$ lies on the shortest path from $p$ to $i$. Consequently, if $d_{jp}\leq d_{ip} / 2$ node $j$ must be at least as close to $p$ as to $i$. Note that $w_{pp'}'$ does not necessarily equal $w_{p'p}'$ for $p,p' \in \P$, and if $k = n$ our model reduces to the full stress model.

\paragraph{Asymptotic running time:} To minimize Eq.~\eqref{eq:sparse} in each iteration we displace all nodes $i \in V$ according to Eq.~\eqref{eq:ssmin}. Since this requires $|N(i)| + k$ constant time operations, given that all graph-theoretic distances are known, the total time per iteration is in \O{kn+m}. Furthermore, only the distances between all $i\in V$ and $p\in \P$ have to be known, which can be done in \O{k(m + n\log n)} time and requires \O{kn} additional space. If the graph-theoretic distances for all $p \in \P$ are computed with a multi-source shortest path algorithm (MSSP), it is possible to construct \R~as well as calculate all $w_{ip}'$ during its execution without increasing its asymptotic running time. The full algorithm to minimize our sparse stress model is presented in \Cref{alg:sparse}.

\section{Experimental Evaluation}\label{sec:ed}

We report on two sets of experiments.
The first is concerned with the evaluation of the impact of different pivot sampling strategies. The second set is designed to assess how well the different sparse stress models approximate the full stress model, in both absolute terms and in relation to the speed-up achieved.

For the experiments we implemented the sparse stress model, \Cref{alg:sparse}, as well as different sampling techniques in Java using Oracle SDK 1.8 and the yFiles 2.9 graph library (\url{www.yworks.com}). The tests were carried out on a single 64-bit machine with a 3.60GHz~quad-core Intel Core~i7-4790 CPU, 32GB~RAM, running Ubuntu~14.10. Times were measured using the \texttt{System.currentTimeMillis()} command. The reported running times were averaged over 25 iterations. We note here that all drawing algorithms, except stated otherwise, were initialized with a 200 \pivmds~layout~\cite{bp-pmds-07}. Furthermore, the maximum number of iterations for the full stress algorithm was set to 500. As stress is not resilient against scaling, see Eq.~\eqref{eq:stress}, we optimally rescaled each drawing such that it creates the lowest possible stress value~\cite{bg-mms-05}.

\begin{table}[t]
	\caption{Dataset: $n$, $m$, $\delta(G)$, $\Delta(G)$, and $D(G)$ denote the number of nodes, edges, the min.~and max. degree, and the diameter, respectively. Column $\{deg(i)\}$ shows the degree and $\{d_{ij}\}$ the distance distribution. Bipartite graphs are marked with $^*$ and weighted graphs with $^{**}$}
		\centering
	 \resizebox{\textwidth}{!}{%
	\begin{tabular}{l@{\enskip}r@{\enskip}r@{\enskip}r@{\enskip}r@{\enskip}r@{\enskip}c@{\enskip}c@{\enskip}||@{\enskip}l@{\enskip}r@{\enskip}r@{\enskip}r@{\enskip}r@{\enskip}r@{\enskip}c@{\enskip}c}
		graph       &   $n$ &   $m$ & $\delta(G)$ & $\Delta(G)$ & $D(G)$  & $\{deg(i)\}$ & $\{d_{ij}\}$ &	graph     &   $n$ &   $m$ & $\delta(G)$ & $\Delta(G)$ & $D(G)$ & $\{deg(i)\}$ & $\{d_{ij}\}$ \\ \hline
		dwt1005     &  1005 &  3808 & 3 & 26 & 34 &  \addDegreeDist{dwt1005}  & \addDistanceDist{dwt1005}  &
		pesa        & 11738 & 33914 & 2 & 9 & 208 &  \addDegreeDist{pesa}   &  \addDistanceDist{pesa}  \\
		1138bus     &  1138 &  1458 & 1 & 17 &  31 & \addDegreeDist{1138bus}   &  \addDistanceDist{1138bus}  &
		bodyy5      & 18589 & 55346 & 2 & 8 & 132 &  \addDegreeDist{bodyy5}   &  \addDistanceDist{bodyy5}  \\
		plat1919    &  1919 & 15240 & 2 & 18 & 43 &  \addDegreeDist{plat1919}   &  \addDistanceDist{plat1919}  &
		finance256  & 20657 & 71866 & 1 & 54 & 55 &  \addDegreeDist{finance256}   &  \addDistanceDist{finance256} \\
		3elt        &  4740 & 13722 & 3 & 9 & 65 &   \addDegreeDist{3elt}  & \addDistanceDist{3elt} &
		btree (binary tree)       &  1023$^*$ &  1022 & 1 & 3 & 18 & \addDegreeDist{btree}  &   \addDistanceDist{btree} \\
		USpowerGrid &  4941 &  6594 & 1 & 19 &  46 &  \addDegreeDist{USpowerGrid}  & \addDistanceDist{USpowerGrid} &
		qh882       &  1764$^*$ &  3354 & 1 & 14 & 32 &  \addDegreeDist{qh882} &  \addDistanceDist{qh882} \\
		commanche   &  7920 & 11880$^{**}$ & 3 & 3 &  438.00 & \addDegreeDist{commanche} &  \addDistanceDist{commanche} &
		lpship04l   &  2526$^*$ &  6380 & 1 & 84 & 13 &  \addDegreeDist{lpship04l} &  \addDistanceDist{lpship04l} \\
		LeHavre     & 11730 & 15133$^{**}$ & 1 & 7 & 33800.67 & \addDegreeDist{LeHavre} &  \addDistanceDist{LeHavre} &&&&&\\
	\end{tabular}}
\label{tab:data}
\end{table}

\paragraph{Data:} We conducted our experiments on a series of different graphs, see \Cref{tab:data}, most of them taken from the sparse matrix collection~\cite{dh-smc-11}. We selected these graphs as they differ in their structure and size, and are large enough to compare the results of different techniques. Two of the graphs, \emph{LeHavre} and \emph{commanche}, have predefined edge lengths that were derived from the node coordinates. We did not modify the graphs in any way, except for those that were disconnected. In this case we only kept the largest component.

\subsection{Sampling Evaluation}
In \Cref{sec:ssm} we discussed how vital the proper selection of the pivots is for our model. In the optimal case we would sample pivots that are well distributed over the graph, creating regions of equal complexity, and are central in the drawing of their regions. In order to evaluate the impact of different sampling strategies on the quality of our sparse stress model and recommend a proper sampling scheme, we compared a set of different strategies:
\begin{itemize}
	\item random: nodes are selected uniformly at random
	\item MIS filtration: nodes are sampled according to the maximal independent set filtration algorithm by Gajer et al.~\cite{gk-grip-00}. Once $n \leq k$ the coarsening stops. If $n < k$, unsampled nodes from the previous level are randomly added
	\item max/min euclidean: starting with a uniformly randomly chosen node, \P~is extended by adding $\argmax_{i \in V \setminus \P} \min_{p \in \P}||x_i-x_p||$
	\item max/min sp: similar to max/min euclidean except that \P~is extended according $\argmax_{i \in V \setminus \P} \min_{p \in \P}d_{ip}$~\cite{bp-pmds-07}
\end{itemize}
Pretests showed that the max/min sp strategy initially favors sampling leaves, but nevertheless produces good results for large $k$. Thus, we also evaluated strategies building on this idea, yet try to overcome the problem of leaf node sampling.
\begin{itemize}
	\item max/min random sp: similar to max/min sp, yet each node $i$ is sampled with a probability proportional to $\min_{p \in \P}d_{ip}$ 
	\item k-means layout: the nodes are selected via a k-means algorithm, running at most 50 iterations, on the initial layout
	\item k-means sp: initially $k$ nodes with max/min sp are sampled succeeded by k-means sampling using the shortest path entries of these pivots
	\item k-means + max/min sp: \P~is initialized with $k/2$ pivots via k-means layout and the remaining nodes are sampled via max/min sp
\end{itemize}

To quantify how well suited each of the sampling techniques is for our model, we ran each combination on each graph with $k \in \{50,51,\dots,200\}$ pivots. For all tests the sparse stress algorithm terminated after 200 iterations. Since all techniques at some point rely on a random decision, we repeated each execution 20 times in order to ensure we do not rest our results upon outliers. To distinguish the applicability of the different techniques to our model, we used two measures. The first measure is the normalized stress, which is the stress value divided by ${n \choose 2}$. While the normalized stress measures the quality of our drawing, we also calculated the Procrustes statistic, which measures how well the layout matches the full stress drawing~\cite{s-proc-78}. The range of the Procrustes statistic is $[0,1]$, where 0 is the optimal match.

The results of these experiments for some of the instances are presented in \Cref{fig:sampleNormStress,fig:sampleProcrustes} (see the Appendix for the full set of data).  
In these plots each dot represents the median and each line starts at the 25\%, 75\% percentile and ends at the 5\%, 95\% percentile, respectively. For the sake of readability we binned each 25 consecutive sample sizes. Furthermore, the strategies were ordered according to their overall ranking w.r.t.~the evaluated measure. For most of the graphs using k-means sp sampling yields the layouts with the lowest normalized stress value. There are only two graphs where this strategy performs worse than other tested strategies. The one graph where k-means sp is outclassed, yet only for large $k$ by max/min sp, is \emph{pesa}. The reason for this result is that k-means sp mainly samples pivots in the center of the left arm, see \Cref{tab:drawings}, creating twists. Max/min sp for small $k$ in contrast mostly samples nodes on the contour of the arm, yet once $k$ reaches a certain threshold the resulting distribution of the pivots prevents twists, yielding a lower normalized stress value.

\begin{figure}[tb]
\centering
\includegraphics[width=.87\textwidth]{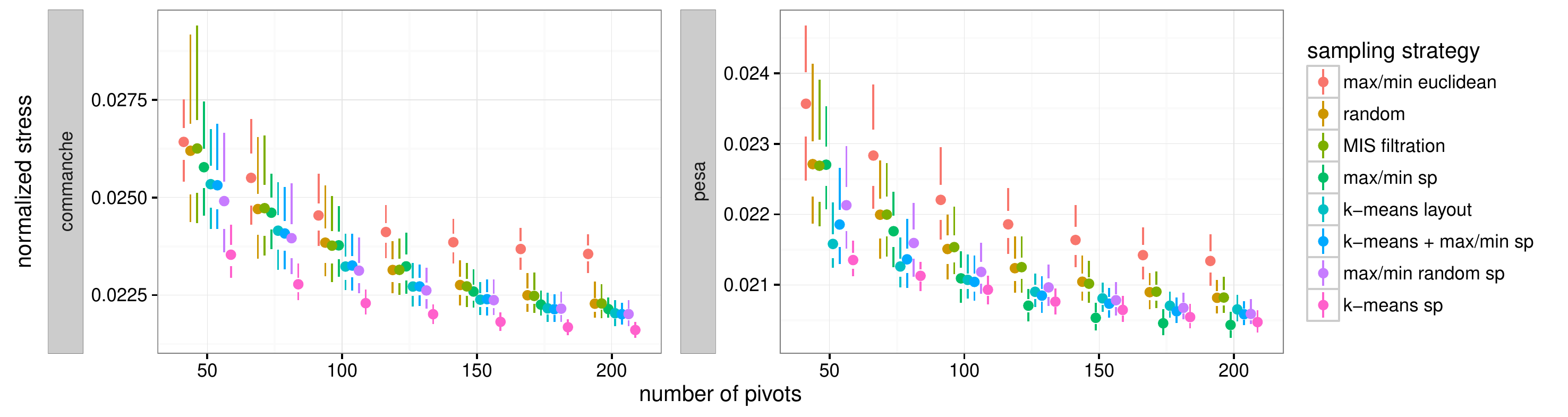}
\caption{Comparison of different sampling strategies and number of pivots w.r.t.~the resulting normalized stress value}
\label{fig:sampleNormStress}
\end{figure}
\begin{figure}[tb]
\centering
\includegraphics[width=.87\textwidth]{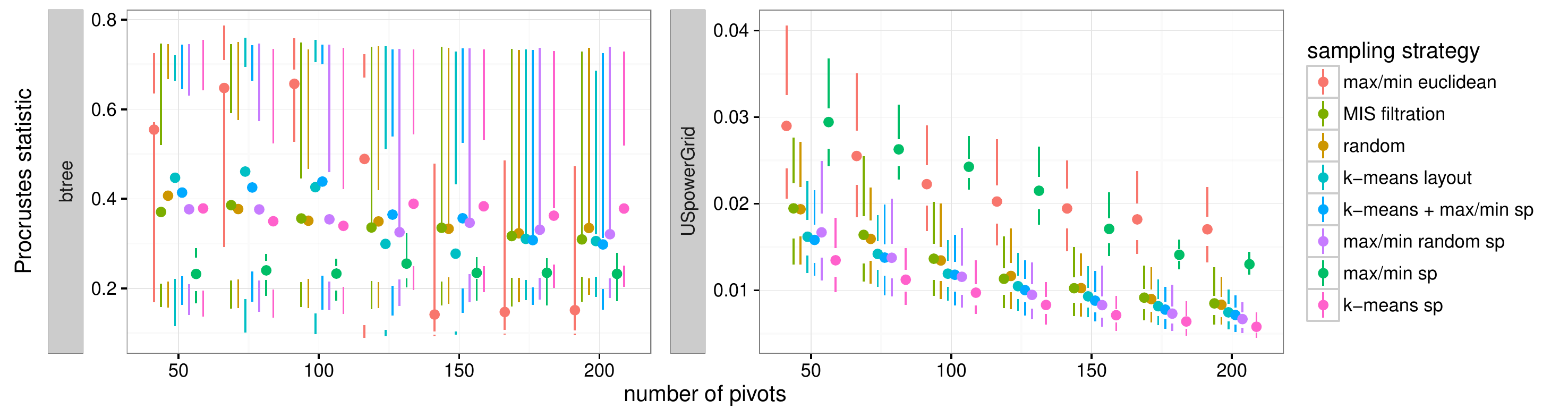}
\caption{Comparison of different sampling strategies and number of pivots w.r.t.~the resulting Procrustes statistic}
\label{fig:sampleProcrustes}
\end{figure}

The explanation of the poor behavior for \emph{lpship04l} is strongly related to its structure. The low diameter of 13 causes, after a few iterations, the max/min sp strategy to repeatedly sample nodes that are part of the same cluster, see \Cref{tab:drawings}, and consequently are structurally very similar. As k-means sp builds on max/min sp, it can only slightly improve the pivot distribution. The argument that the problem is related to the structure is reinforced by the outcome of the random strategy. Still, except for these two graphs k-means sp generates the best outcomes, and since this strategy is also strongly favorable over the others subject to the Procrustes statistics, see \Cref{fig:sampleProcrustes}, our following evaluation always relies on this sampling strategy. 
However, we note that the Procrustes statistic for \emph{btree} and \emph{lpship04l} are by magnitudes larger than for any other tested graph. While for \emph{lpship04l} this is mostly caused by the quality of the drawings, this is only partly true for \emph{btree}. 
The other factor contributing to the high Procrustes statistic for \emph{btree} is caused by the restricted set of operations provided by the Procrustes analysis. As dilation, translation, and rotation are used to find the best match between two layouts, the Procrustes analysis cannot resolve reflections. Therefore, if in the one layout of \emph{btree}, the subtree $T_1$ of $v$ is drawn to the right of subtree $T_2$ of $v$ and vice versa in the second drawing, although the two layouts are identical, the statistic will be high. This symmetry problem mainly explains the low performance w.r.t.~\emph{btree}.

\begin{table}[tb]
\caption{Stress and Procrustes statistics: sparse model values are highlighted when no larger than minimum over previous methods}
\label{tab:stress_proc}
\centering
\resizebox{0.89\textwidth}{!}{%
	\begin{tabular}{l@{\quad}r@{\quad}||@{\quad}r@{\quad}r@{\quad}r@{\quad}|@{\quad}r@{\quad}r@{\quad}r@{\quad}r@{\quad}r@{\quad}r}
		graph       & \stress   & \sparsetwo & \sparseone & \sparsefif &    \maxent &   \marstwo &   \marsone &      \grip &        \bp &    \pivmds \\ \hline
		\multicolumn{11}{c}{\cellcolor{gray!40}\textbf{stress}}\\\hline
		dwt1005     & 10\,729    &     \textbf{10\,940} &     \textbf{11\,081} &     \textbf{11\,329} &     21\,623 &     17\,660 &     20\,134 &     52\,517 &     \colorbox{gray!40}{12\,495} &     14\,459 \\
		 1138bus     & 39\,974    &     \textbf{40\,797} &     \textbf{41\,471} &     \textbf{42\,686} &     \colorbox{gray!40}{44\,650} &     64\,363 &     63\,614 &     54\,986 &     73\,512 &     75\,427 \\
		plat1919    & 18\,572    &     \textbf{18\,840} &     \textbf{19\,072} &     \textbf{19\,719} &     \colorbox{gray!40}{23\,850} &     53\,246 &     64\,166 &    113\,765 &     75\,973 &     82\,865 \\
		3elt        & 422\,940  &    \textbf{426\,564} &    \textbf{430\,200} &    \textbf{437\,051} &    585\,967 &    \colorbox{gray!40}{503\,600} &    754\,134 &    934\,206 &    555\,934 &    634\,401 \\
		USpowerGrid & 702\,055   &    \textbf{720\,642} &    \textbf{731\,187} &    \textbf{749\,464} &  1\,021\,457 &    \colorbox{gray!40}{766\,535} &    783\,888 &  1\,495\,373 &  1\,111\,216 &  1\,123\,698 \\
		commanche   & 654\,694   &    \textbf{677\,220} &    \textbf{699\,890} &    \textbf{749\,609} &  \colorbox{gray!40}{1\,507\,654} &  2\,761\,605 &  3\,145\,489 &  1\,539\,767 &  2\,085\,818 &  2\,157\,943 \\
		LeHavre     & 439\,188   &    \textbf{433\,030} &    \textbf{441\,986} &    \textbf{454\,785} &  \colorbox{gray!40}{1\,231\,283} & 12\,012\,307 & 12\,570\,692 &  8\,658\,371 &  1\,255\,474 &  1\,305\,577 \\
		pesa        & 1\,373\,514 &  \textbf{1\,417\,449} &  \textbf{1\,452\,975} &  \textbf{1\,495\,512} & 10\,423\,779 &  3\,563\,772 &  8\,281\,116 &  \colorbox{gray!40}{2\,957\,738} &  3\,486\,176 &  3\,325\,889 \\
		bodyy5      & 3\,547\,659 &  \textbf{3\,566\,636} &  \textbf{3\,585\,087} &  \textbf{3\,630\,380} &  5\,248\,755 &  6\,385\,559 &  \colorbox{gray!40}{4\,072\,905} & 10\,389\,846 &  4\,245\,006 &  4\,715\,728 \\
		finance256  & 6\,175\,210 &  \textbf{6\,415\,761} &  \textbf{6\,474\,787} &  \textbf{6\,582\,890} &  8\,151\,335 &  \colorbox{gray!40}{7\,267\,598} &  8\,643\,239 & 19\,817\,355 & 12\,257\,268 & 11\,380\,089 \\
		btree       & 60\,206    &     \textbf{61\,839} &     \textbf{63\,325} &     \textbf{66\,122} &     \colorbox{gray!40}{67\,871} &    103\,436 &    100\,767 &     96\,235 &    157\,988 &    164\,329 \\
		qh882       & 84\,524    &     \textbf{86\,345} &     \textbf{87\,695} &     \textbf{89\,556} &   \colorbox{gray!40}{103\,601} &    117\,195 &    161\,113 &    127\,914 &    146\,935 &    143\,142 \\
		lpship04l   & 250\,599 &    \textbf{297\,547} &    \textbf{316\,674} &    343\,694 &    \colorbox{gray!40}{329\,255} &    558\,923 &    542\,667 &    771\,284 &    775\,813 &    793\,238 \\\hline
		\multicolumn{11}{c}{\cellcolor{gray!40}\textbf{Procrustes statistic}}\\\hline
		dwt1005 & & \textbf{0.001} & \textbf{0.005} & \textbf{0.003} & 0.027 & \colorbox{gray!40}{0.008} & 0.018 & 0.263 & 0.004 & \colorbox{gray!40}{0.008}\\
		1138bus &  &\textbf{0.009} & \textbf{0.016} & 0.025 & \colorbox{gray!40}{0.022} & 0.148 & 0.145 & 0.071 & 0.097 & 0.102\\
		plat1919 & & \textbf{0.000} & \textbf{0.000} & \textbf{0.001} & \colorbox{gray!40}{0.015} & 0.026 & 0.031 & 0.236 & 0.045 & 0.051\\
		3elt &  &\textbf{0.001} & \textbf{0.001} & \textbf{0.002} & 0.026 & \colorbox{gray!40}{0.009} & 0.029 & 0.199 & 0.017 & 0.023\\
		USpowerGrid &  &\textbf{0.006} & \textbf{0.008} & \textbf{0.012} & 0.068 & \colorbox{gray!40}{0.014} & 0.018 & 0.256 & 0.051 & 0.051\\
		commanche & & \textbf{0.001} & \textbf{0.002} & \textbf{0.005} & 0.039 & \colorbox{gray!40}{0.026} & 0.167 & 0.092 & 0.066 & 0.066\\
		LeHavre & & \textbf{0.001} & \textbf{0.001} & \textbf{0.001} & 0.012 & 0.163 & 0.173 & 0.256 & \colorbox{gray!40}{0.010} & \colorbox{gray!40}{0.010}\\
		pesa &  &\textbf{0.009} & \textbf{0.010} & \textbf{0.010} & 0.095 & 0.025 & 0.070 & \colorbox{gray!40}{0.017} & 0.021 & 0.021\\
		bodyy5 &  &\textbf{0.000} & \textbf{0.000} & \textbf{0.000} & 0.012 & 0.011 & \colorbox{gray!40}{0.003} & 0.100 & 0.004 & 0.007\\
		finance256 &  &0.009 & \textbf{0.006} & \textbf{0.005} & 0.013 & \colorbox{gray!40}{0.007} & 0.018 & 0.206 & 0.042 & 0.041\\
		btree & & 0.748 & \textbf{0.165} & 0.241 & \colorbox{gray!40}{0.233} & 0.360 & 0.367 & 0.386 & 0.361 & 0.364\\
		qh882 & &\textbf{0.015} & \textbf{0.015} & \textbf{0.021} & \colorbox{gray!40}{0.046} & 0.061 & 0.114 & 0.075 & 0.086 & 0.079\\
		lpship04l & & 0.176 & \textbf{0.112} & \textbf{0.148} & \colorbox{gray!40}{0.160} & 0.246 & 0.587 & 0.463 & 0.393 & 0.401
	\end{tabular}}
\end{table}
\begin{table}[tb]
\caption{Runtime in seconds: fastest sparse model yielding lower stress than best previous method, c.f.~\autoref{tab:stress_proc}, is highlighted. Marked implementations written in \texttt{C/C++} with time measured via \texttt{clock()} command}
\label{tab:time}
\centering
\resizebox{0.89\textwidth}{!}{%
	\begin{tabular}{l@{\quad}r@{\quad}||@{\quad}r@{\quad}r@{\quad}r@{\quad}|@{\quad}r@{\quad}r@{\quad}r@{\quad}r@{\quad}r@{\quad}r}
		graph       & \stress & \sparsetwo  & \sparseone & \sparsefif & \maxent$^*$ & \marstwo$^*$ & \marsone$^*$ & \grip$^*$ &  \bp & \pivmds \\ \hline
		dwt1005     &    1.26 & 0.33 & 0.15 & \textbf{0.09} &  0.47 &  1.02 &  2.36 &  0.06 & \colorbox{gray!40}{0.08} & 0.06\\
		1138bus     &    2.20 & 0.41 & 0.16 & \textbf{0.09} &  \colorbox{gray!40}{0.91} &  3.16 &  1.96 &  0.20 & 0.06 & 0.04\\
		plat1919    &   9.70 & 1.00 & 0.45 & \textbf{0.24} &   \colorbox{gray!40}{1.15} &  6.80 &  4.79 &  0.19 & 0.31 & 0.20\\
		3elt        &   31.82 & 2.28 & 0.93 & \textbf{0.43} &  2.26 & \colorbox{gray!40}{16.31} &  8.43 &  0.71 & 0.37 & 0.23\\
		USpowerGrid &   36.48 & 1.85 & 0.67 & \textbf{0.37} &  2.53 & \colorbox{gray!40}{13.54} &  7.62 &  1.67 & 0.28 & 0.21\\
		commanche   &  340.10 & 10.78 & 3.63 & \textbf{1.51} &  \colorbox{gray!40}{3.60} & 22.72 & 12.43 &  2.29 & 0.47 & 0.35\\
		LeHavre     &  475.05 & 12.75 & 4.90 & \textbf{2.19} &  \colorbox{gray!40}{6.31} & 27.57 & 19.50 & 10.18 & 0.81 & 0.54\\
		pesa        &  373.23 & 9.61 & 4.14 & \textbf{1.50} &  5.96 & 50.10 & 42.68 & \colorbox{gray!40}{3.56} & 0.95 & 0.60\\
		bodyy5      & 463.47 & 12.53 & 4.31 & \textbf{2.01} &   9.97 & 46.63 & \colorbox{gray!40}{9.27} & 10.43 & 1.64 & 1.04\\
		finance256  & 1016.92 & 10.44 & 4.27 & \textbf{2.28} & 14.76 & \colorbox{gray!40}{32.16} & 24.66 & 12.12 & 2.51 & 1.60\\
		btree       &    7.79 & 0.42 & 0.18 & \textbf{0.09} & \colorbox{gray!40}{0.63} & 2.70 & 1.48 & 0.06 & 0.06 & 0.03\\
		qh882       &    6.61 & 0.65 & 0.28 & \textbf{0.15} &  \colorbox{gray!40}{0.97} & 8.45 & 5.79 & 0.15 & 0.17 & 0.14\\
		lpship04l   &   18.30 & 0.73 & \textbf{0.31} & 0.18 &  \colorbox{gray!40}{0.99} & 7.06 & 7.63 & 0.16 & 0.15 & 0.10
	\end{tabular}}
\end{table}

\begin{figure}[tb]
\centering
	\includegraphics[width=.87\textwidth]{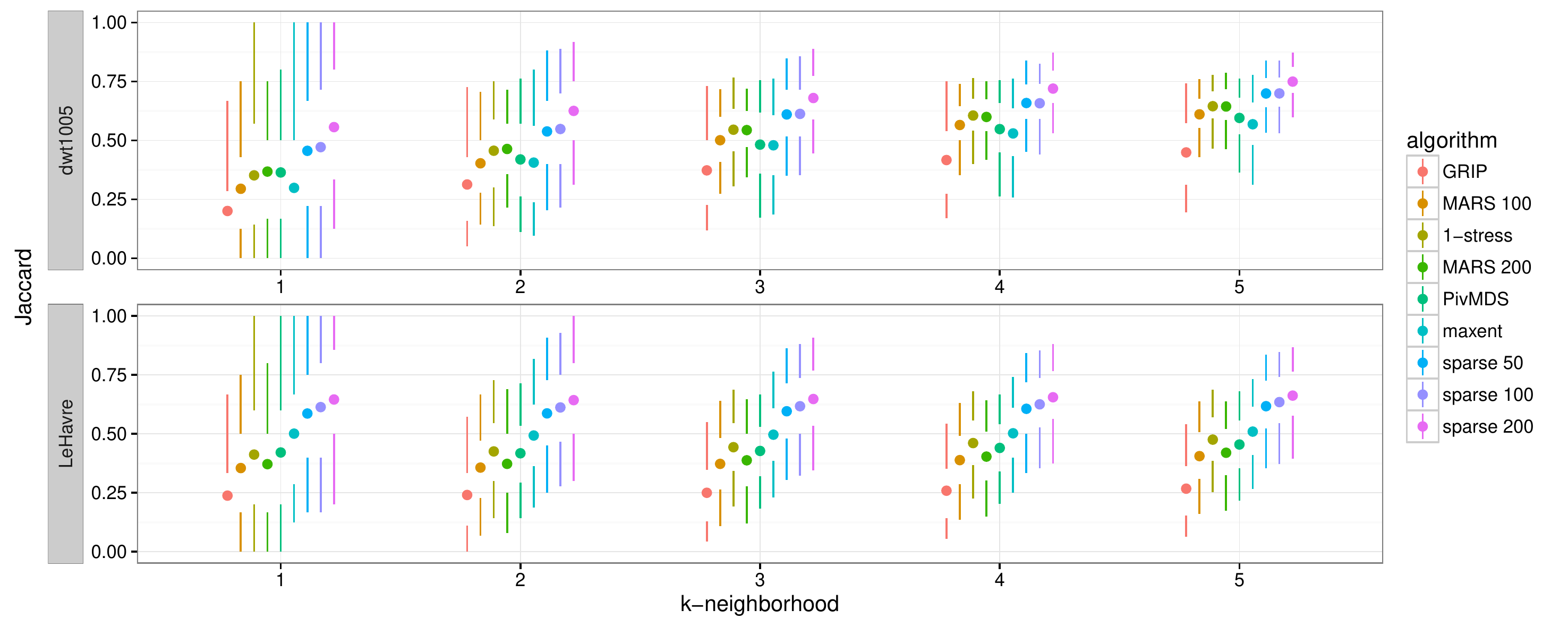}
	\caption{The similarity of the Gabriel Graph of the full stress layout and the Gabriel Graph of the layout algorithms under consideration as a function of $k$. For each node of the graph the k-neighborhood in the Gabriel Graph of the full stress layout and the layout algorithm are compared by calculating the Jaccard coefficient. A higher value indicates that the nodes share a high percentage of common neighbors in the different Gabriel Graphs.}
	\label{fig:gg}
\end{figure}
\begin{figure}[tb]
\centering
	\includegraphics[width=.87\textwidth]{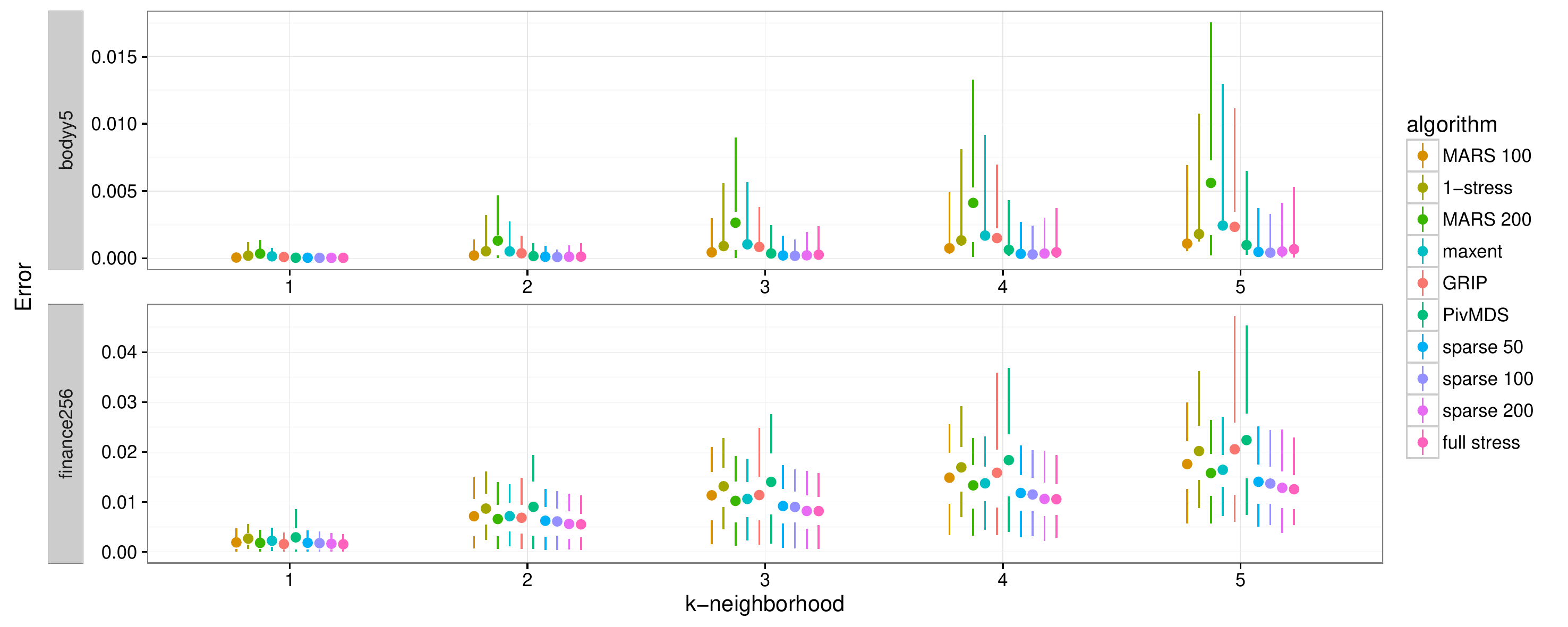}
	\caption{Error charts as a function of $k$. For each node of the graph the convex hull w.r.t.~the coordinates of the nodes in the k-neighborhood is computed. For each of the convex hulls the error is calculated by counting the number of non k-neighborhood nodes that lie inside or on the contour of this hull divided by their total number.}
	\label{fig:ch}
\end{figure}

\subsection{Full Stress Layout Approximation}
The next set of experiments is designed to assess how well our sparse stress model using k-means sp sampling, as well as related sparse stress techniques, resembles the full stress model. For this we compared the median stress layout over 25 repetitions on the same graph of our sparse stress model with $k \in \{50,100,200\}$, with \mars,\footnote{\url{https://github.com/marckhoury/mars}} \maxent,\footnote{We are grateful to Yifan Hu for providing us with the code.} \pivmds, \bp, and the weighted version of \grip.\footnote{\url{http://www.cs.arizona.edu/~kobourov/GRIP/}} The number of iterations of our model as well as for \mars~and \bp~have been limited to 200. Furthermore, we tested \mars~with 100 and 200 pivots and report the layout with the smallest stress from the drawings obtained by running \texttt{mars} with argument \texttt{-p} $\in \{1,2\}$ combined with a \pivmds~or randomly initialized layout.

Besides comparing the resulting stress values and Procrustes statistics, we compared the distribution of pairwise euclidean distances subject to their graph-theoretic distances. Since the Procrustes statistic has problems with symmetries, as we pointed out in the previous subsection, we propose to evaluate the similarity of the sparse stress layouts with the \stress~layout via Gabriel graphs~\cite{gs-gg-69}. The Gabriel graph of a given layout $\layVar$ contains an edge between a pair of points if and only if the disc associated with the diameter of the endpoints does not contain any other point. Since the treatment of identical positions is not defined for Gabriel Graphs, we resolve this by adding edges between each pair of identical positions. We assess the similarity between the Gabriel Graph of the \stress~layout and the sparse stress layouts by comparing the k-neighborhoods of a node in the graphs using the Jaccard coefficient.

A further measure we introduce evaluates the visual error. More precisely we measure for a given node $v$ the percentage of nodes that lie in the drawing area of the k-neighborhood, but are not part of it. We calculate this value by computing the convex hull induced by the k-neighborhood and then test for each other node if it belongs to the hull or not. This number is then divided by $n - |\{w \in V | d_{vw} \leq k\}|$. Therefore, a low value implies that there are only a few nodes lying in the region, while high values indicate we cannot distinguish non k-neighborhood and k-neighborhood nodes in the drawing. This measure is to a certain extend similar to the precision of neighborhood preservation~\cite{ghn-msm-13}.

The results of all these experiments, see \Cref{tab:stress_proc,tab:drawings}, \Cref{fig:gg,fig:ch}, and the Appendix, reveal that our model is more adequate in resembling the full stress drawing than any other of the tested algorithm, while showing comparable running times that scale nicely with $k$, cf.~\Cref{tab:time}. The error plots in \Cref{tab:drawings} expose the strength of our approximation scheme. We can see that, while all approaches work very well in representing short distances, our approach is more precise in approximating middle and especially long distances, explaining our good results. As the evaluation clearly shows that our approach yields better approximations of the full stress model, we rather want to discuss the low performance of our model for \emph{lpship04l} and thereby expose one weakness of our approach.

Looking at the sparse 50 drawing of \emph{lpship04l} in \Cref{tab:drawings}, we can see that a large portion of nodes share a similar or even the same position. This is because \emph{lpship04l} has a lot of nodes that share very similar graph-theoretic distance vectors, exhibit highly overlapping neighborhoods, and are drawn in close proximity in the initial \pivmds~layout. While our model would rely on small variations of the graph-theoretic distances to create a good drawing we diminish these differences even further by restricting our model to \P. Consequently, the positional vote for two similar non-pivot nodes $i$ and $j$ that lie in the same partition will only slightly differ, mainly caused by their distinct neighbors. However, as these neighbors are also in close proximity in the initial drawing of \emph{lpship04l} the distance between $i$ and $j$ will not increase. Therefore, if the graph has a lot of structurally very similar nodes and the initial layout has poor quality, our approach will inevitably create drawings where nodes are placed very close to one another.

\section{Conclusion}\label{sec:con}
In this paper we proposed a sparse stress model that requires \O{kn+m} space and time per iteration, and a preprocessing time of \O{k(m+n\log n)}. While Barnes \& Hut derive their representatives from a given partitioning, we argued that for our model it is more appropriate to first select the pivots and then to partition the graph only relying on its structure. Since the approximation quality heavily depends on the proper selection of these pivots, we evaluated different sampling techniques, showing that k-means sp works very well in practice. 

Furthermore, we compared a variety of sparse stress models w.r.t.~their
performance in approximating the full stress model. We therefore proposed two
new measures to assemble the similarity between two layouts of the same graph. For
the tested graphs, all our experiments clearly showed that our proposed sparse
stress model exceeds related approaches in approximating the full stress layout
without compromising the computation time.

\begin{table}[tb]
\caption{Layouts and error charts of the algorithms. Each chart shows the zero y coordinate (black horizontal line), the median (red line), the 25 and 75 percentiles (black/gray ribbon) and the min/max error (outer black dashed line). The error (y-axis) is the difference between the euclidean distance and the graph-theoretic distance (x-axis). 1000 bins have been used for weighted graphs}
\label{tab:drawings}
\centering
\resizebox{\textwidth}{!}{%
 \renewcommand{\arraystretch}{0.4}
\begin{tabular}{cc||ccc|cccccc}
\LARGE	graph  & \LARGE	\stress & \LARGE	\sparsetwo & \LARGE	\sparseone & \LARGE	\sparsefif &    \LARGE	\maxent &   \LARGE	\marstwo &   \LARGE	\marsone &      \LARGE	\grip &        \LARGE	\bp &    \LARGE	\pivmds   \\\hline &&&&&&&&&&\\
\addLayoutAndOverviewToTable{dwt1005}\\
\addLayoutAndOverviewToTable{1138bus}\\
\addLayoutAndOverviewToTable{plat1919}\\
\addLayoutAndOverviewToTable{3elt}\\
\addLayoutAndOverviewToTable{commanche}\\
\addLayoutAndOverviewToTable{pesa}\\
\addLayoutAndOverviewToTable{finance256}\\
\addLayoutAndOverviewToTable{qh882}\\
\addLayoutAndOverviewToTable{lpship04l}\\
\end{tabular}}

\end{table}

\bibliographystyle{splncs03}

\bibliography{references,../../bib/strings,../../bib/references,../../bib/algo}

\newpage

\appendix
\section*{Appendix}\label{sec:app}

\begin{table}[h]
\caption{Layouts and error charts of the algorithms. Each chart shows the zero y coordinate (black horizontal line), the median (red line), the 25 and 75 percentiles (black/gray ribbon) and the min/max error (outer black dashed line). The error (y-axis) is the difference between the euclidean distance and the graph-theoretic distance (x-axis). 1000 bins have been used for weighted graphs}
\centering
\resizebox{\textwidth}{!}{%
 \renewcommand{\arraystretch}{0.4}
\begin{tabular}{cc||ccc|cccccc}
\LARGE	graph  & \LARGE	\stress & \LARGE	\sparsetwo & \LARGE	\sparseone & \LARGE	\sparsefif &    \LARGE	\maxent &   \LARGE	\marstwo &   \LARGE	\marsone &      \LARGE	\grip &        \LARGE	\bp &    \LARGE	\pivmds   \\\hline &&&&&&&&&&\\
\addLayoutAndOverviewToTable{dwt1005}\\
\addLayoutAndOverviewToTable{1138bus}\\
\addLayoutAndOverviewToTable{plat1919}\\
\addLayoutAndOverviewToTable{3elt}\\
\addLayoutAndOverviewToTable{USpowerGrid}\\
\end{tabular}}
\end{table}

\begin{table}
\centering
\resizebox{\textwidth}{!}{%
 \renewcommand{\arraystretch}{0.4}
\begin{tabular}{cc||ccc|cccccc}
\LARGE	graph  & \LARGE	\stress & \LARGE	\sparsetwo & \LARGE	\sparseone & \LARGE	\sparsefif &    \LARGE	\maxent &   \LARGE	\marstwo &   \LARGE	\marsone &      \LARGE	\grip &        \LARGE	\bp &    \LARGE	\pivmds   \\\hline &&&&&&&&&&\\
\addLayoutAndOverviewToTable{commanche}\\
\addLayoutAndOverviewToTable{LeHavre}\\
\addLayoutAndOverviewToTable{pesa}\\
\addLayoutAndOverviewToTable{bodyy5}\\
\addLayoutAndOverviewToTable{finance256}\\
\addLayoutAndOverviewToTable{btree}\\
\addLayoutAndOverviewToTable{qh882}\\
\addLayoutAndOverviewToTable{lpship04l}\\
\end{tabular}}
\end{table}

\begin{figure}
\centering
	\includegraphics[width=.9\textwidth,height= 3cm]{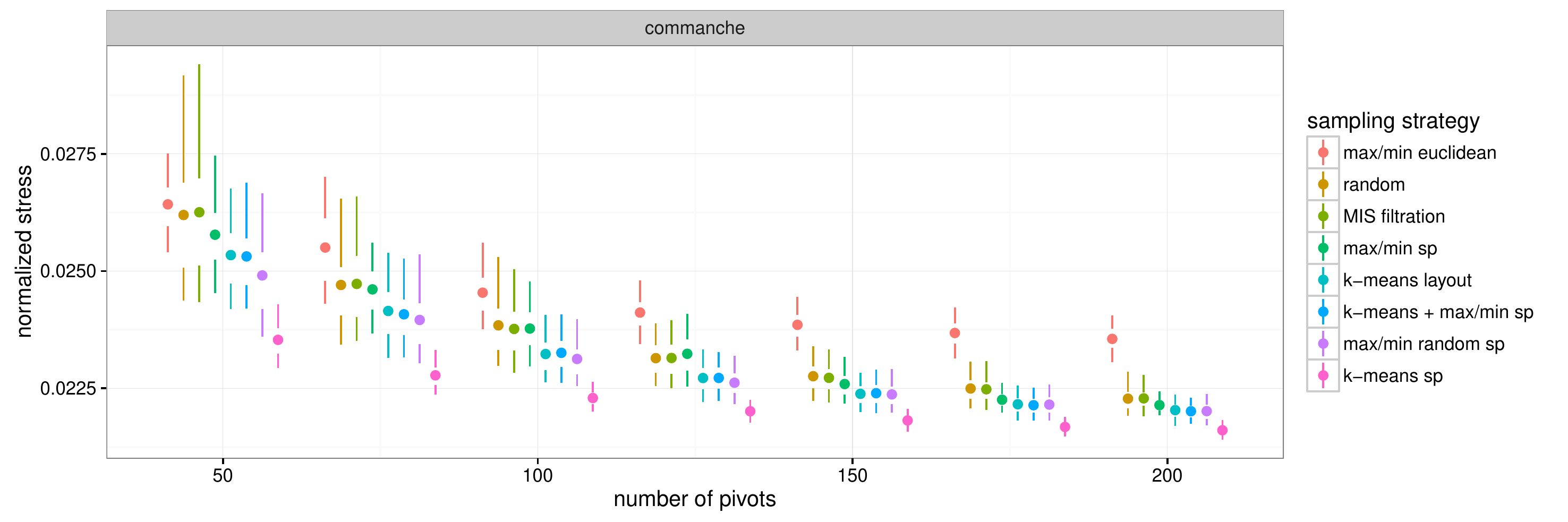}
	\vspace{0.5cm}
	\includegraphics[angle = 90, width=.9\textwidth]{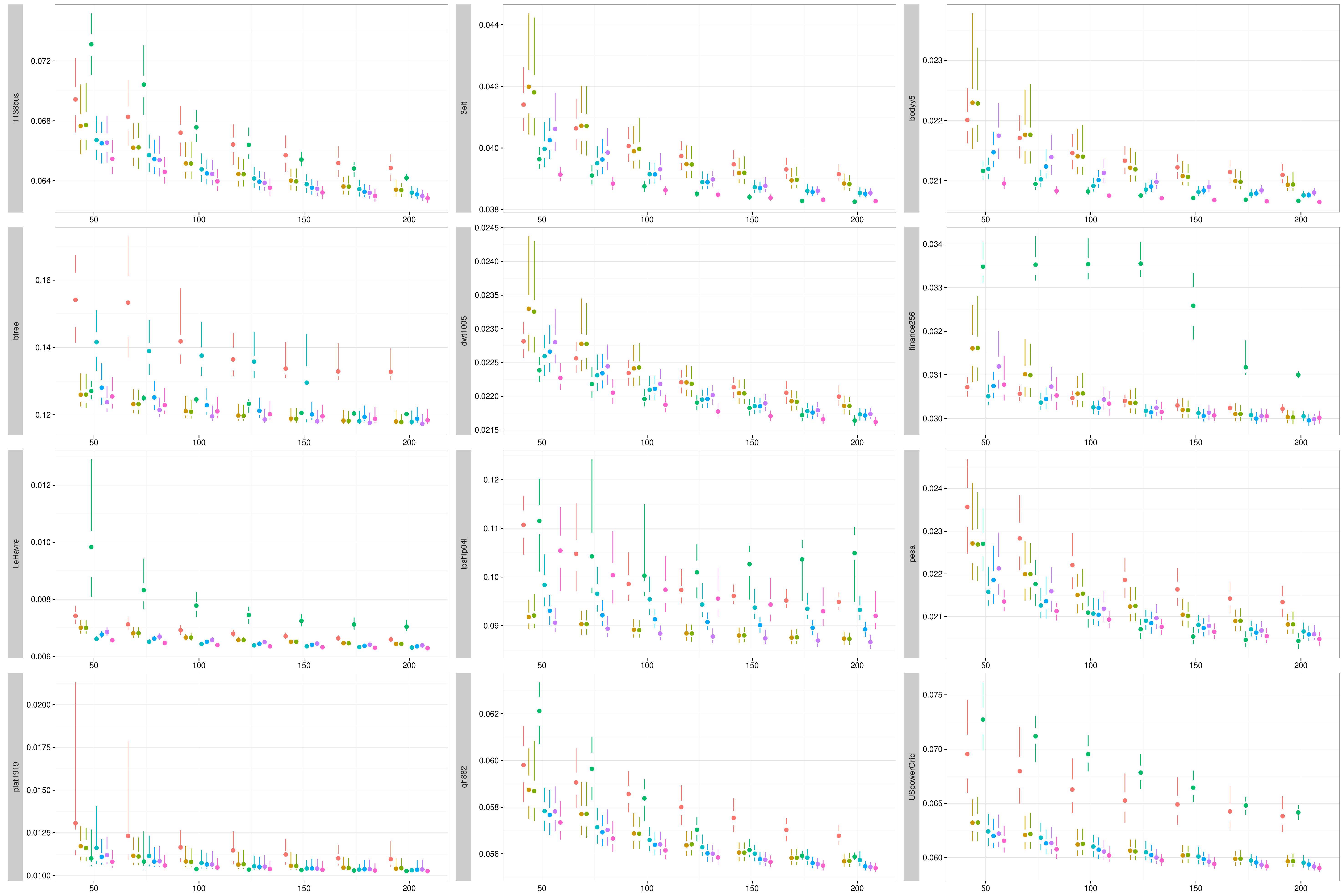}
	\caption{Comparison of different sampling strategies and number of pivots w.r.t.~the resulting normalized stress value}
\end{figure}

\begin{figure}
\centering
	\includegraphics[width=.9\textwidth,height= 3cm]{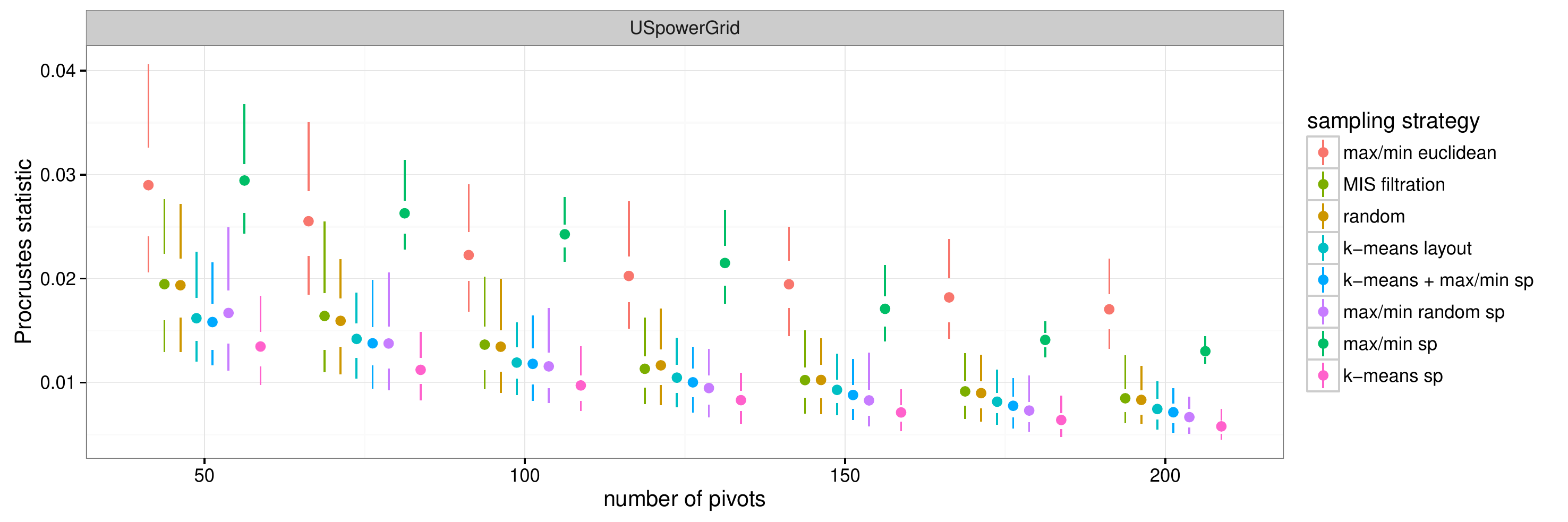}
	\vspace{0.5cm}
	\includegraphics[angle = 90, width=.9\textwidth]{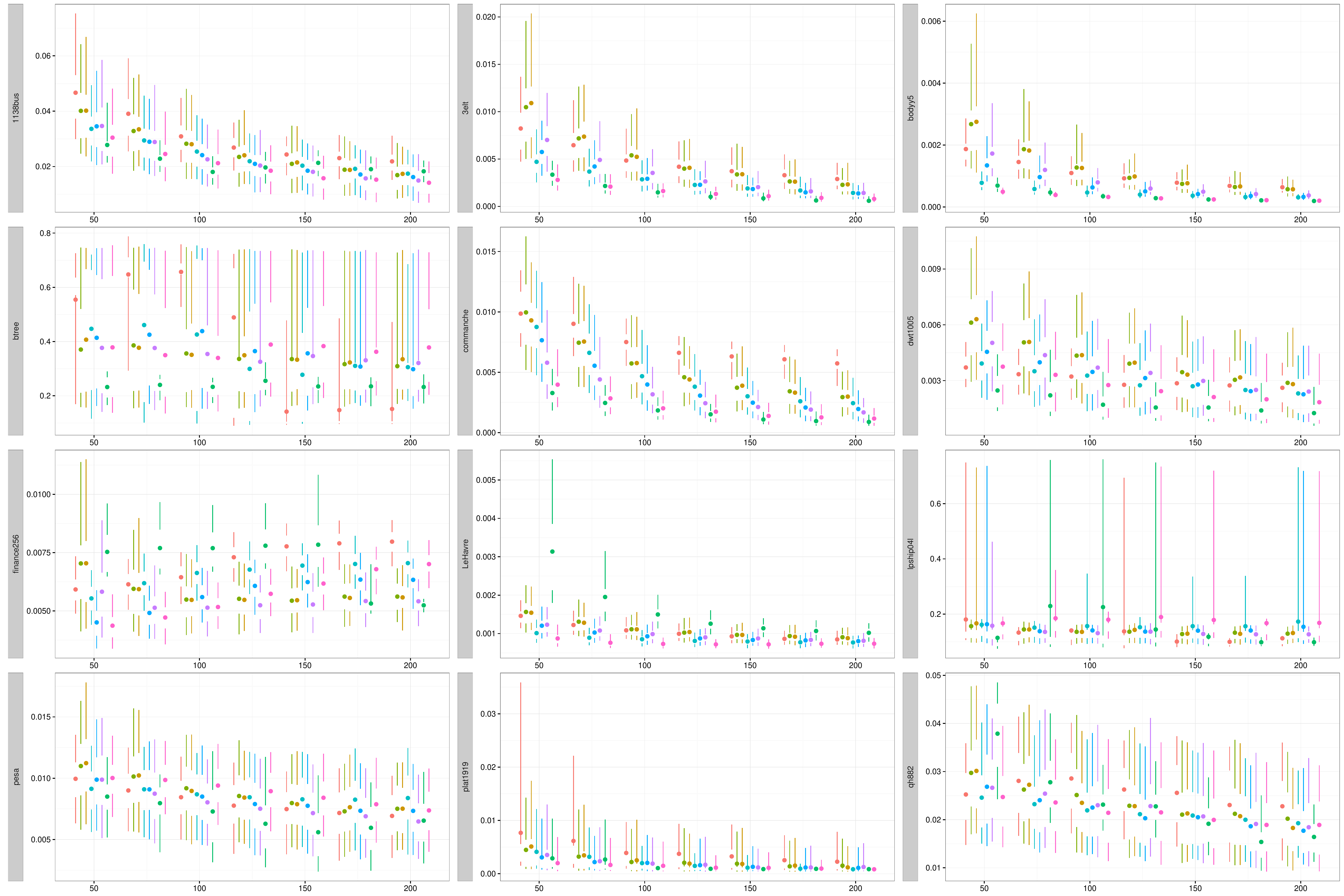}
	\caption{Comparison of different sampling strategies and number of pivots w.r.t.~the Procrustes statistic}
\end{figure}

\begin{figure}
\centering
	\includegraphics[width=.9\textwidth,height= 3cm]{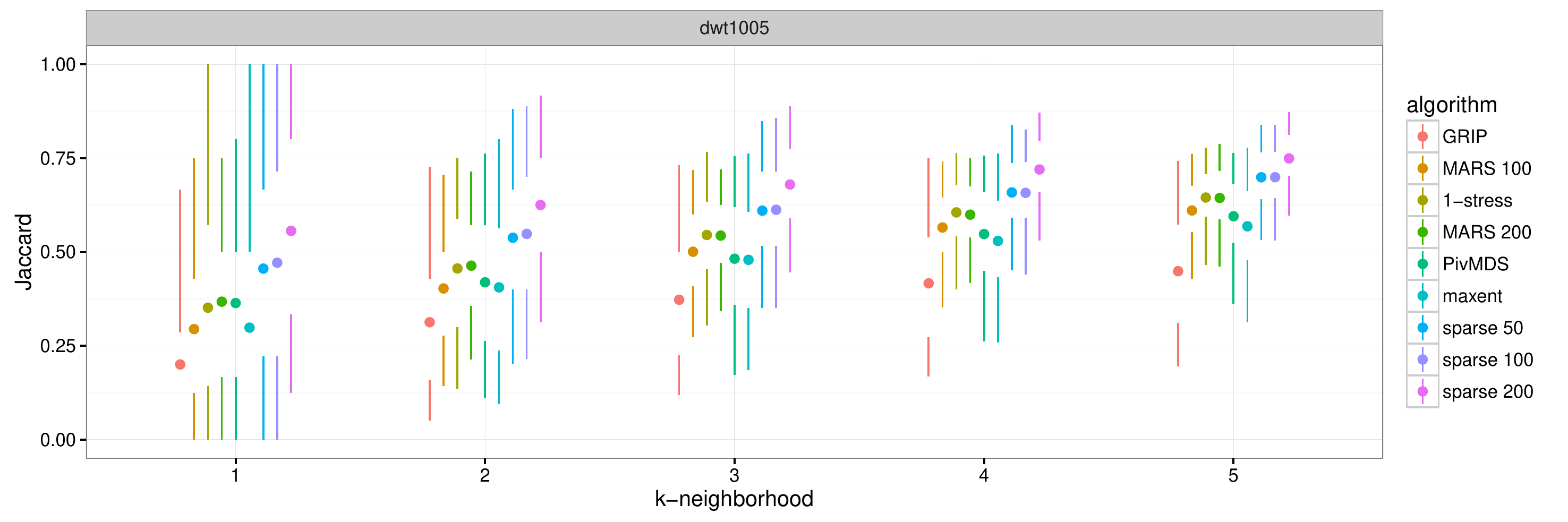}
	\vspace{0.5cm}
	\includegraphics[angle = 90, width=.9\textwidth]{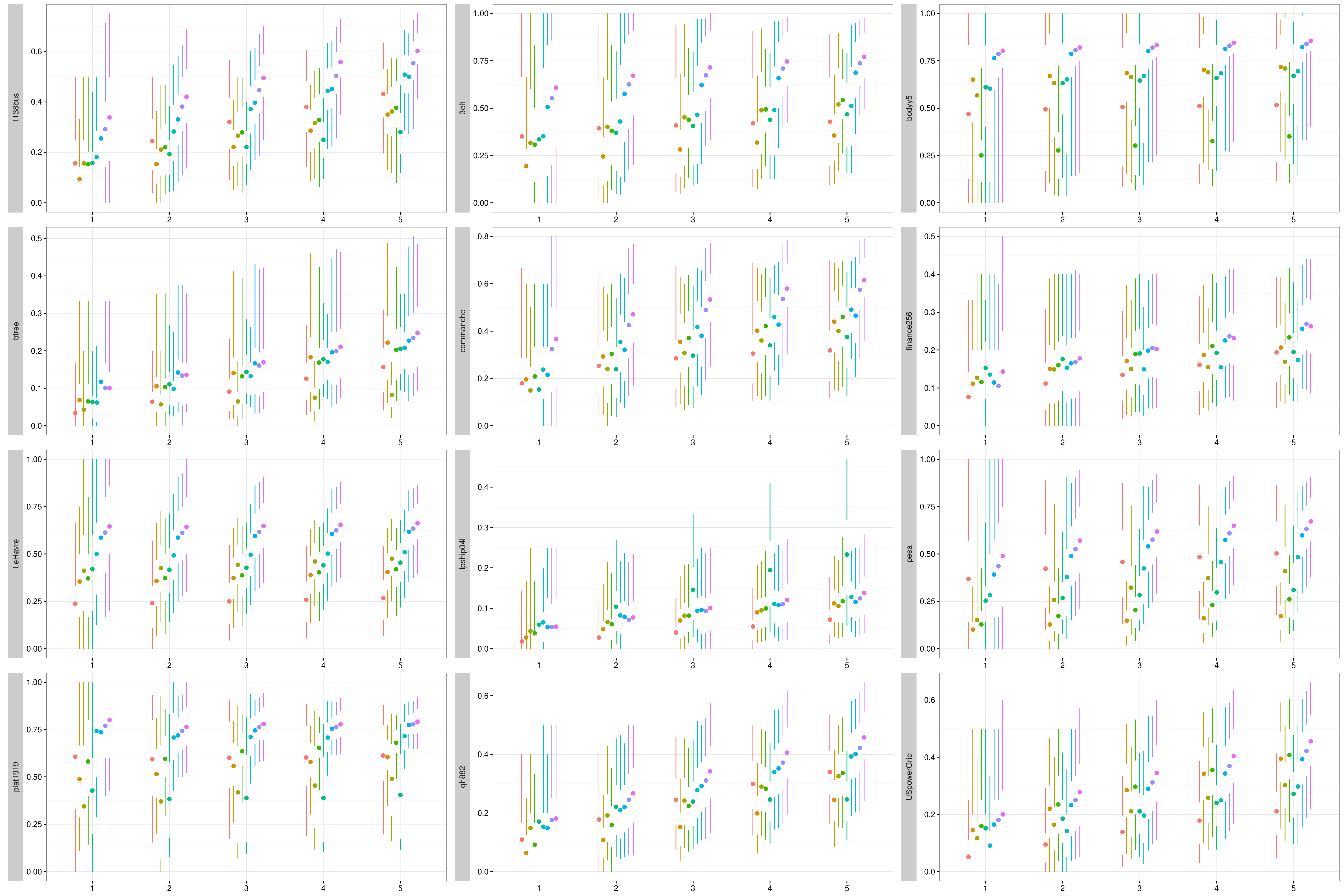}
	\caption{The similarity of the Gabriel Graph of the full stress layout and the Gabriel Graph of the layout algorithms under consideration as a function of $k$. For each node of the graph the k-neighborhood in the Gabriel Graph of the full stress layout and the layout algorithm are compared by calculating the Jaccard coefficient. A higher value indicates that the nodes share a high percentage of common neighbors in the different Gabriel Graphs.}
\end{figure}

\begin{figure}
\centering
	\includegraphics[width=.9\textwidth,height= 3cm]{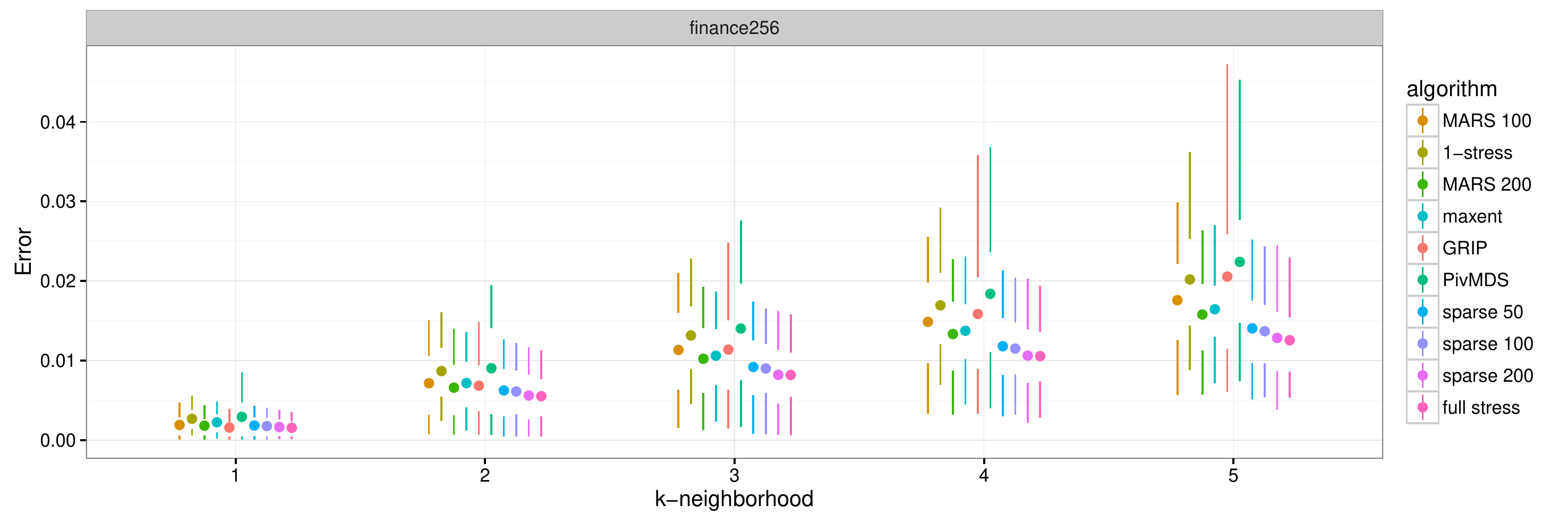}
	\vspace{0.5cm}
	\includegraphics[angle = 90, width=.9\textwidth]{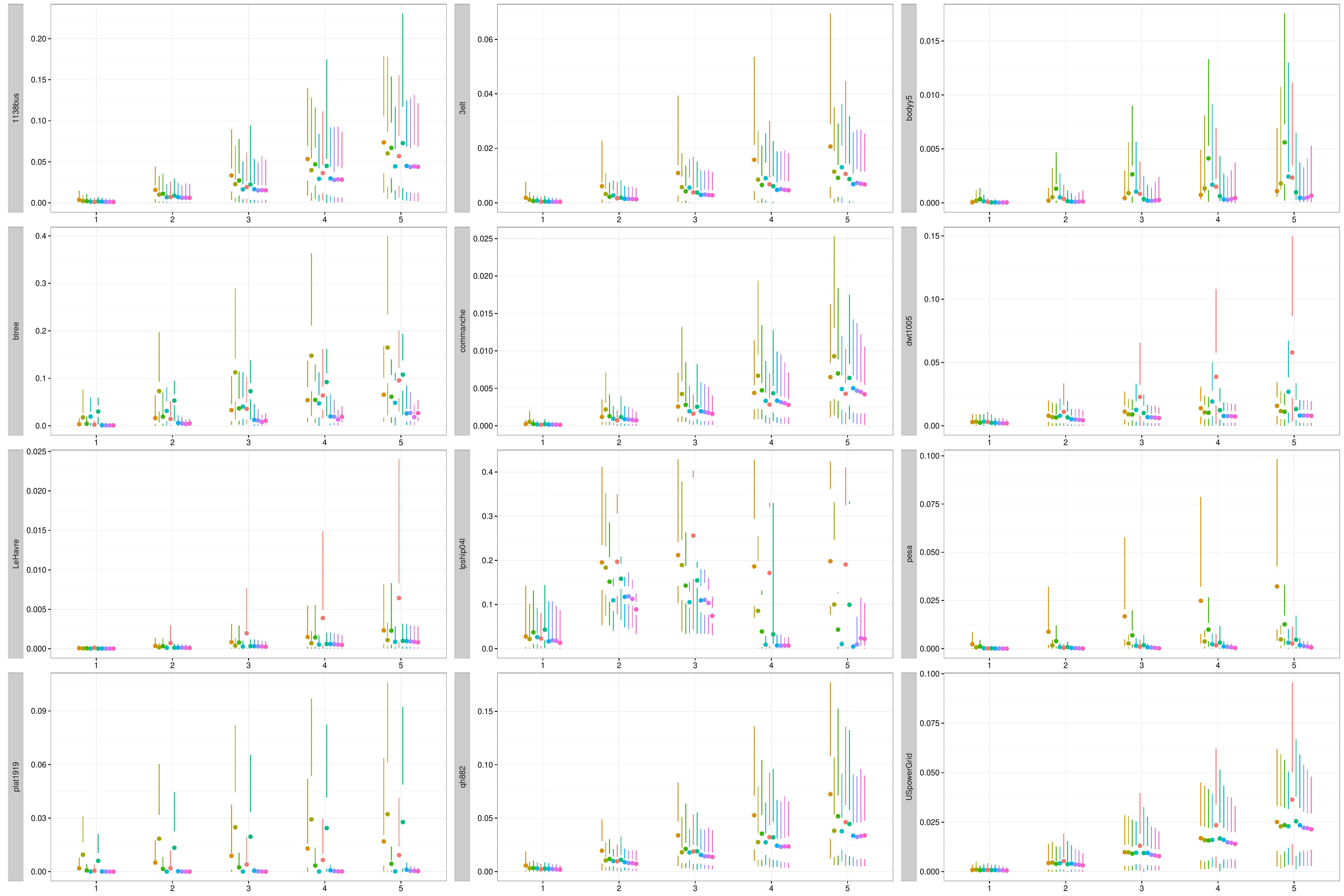}
	\caption{Error charts as a function of $k$. For each node of the graph the convex hull w.r.t.~the coordinates of the nodes in the k-neighborhood is computed. For each of the convex hulls the error is calculated by counting the number of non k-neighborhood nodes that lie inside or on the contour of this hull divided by their total number.}
\end{figure}

\end{document}